\documentclass[letterpaper,A4]{article}

\usepackage{osameet3} 

\usepackage{amsmath}
\usepackage{amssymb}
\usepackage{mathbbol}
\usepackage[pdftex,colorlinks=true,bookmarks=false,citecolor=blue,urlcolor=blue]{hyperref} 

\newcommand{\beqa}{\begin{eqnarray}}
\newcommand{\eeqa}{\end{eqnarray}}

\newcommand{\ket}[1]{\left| #1 \right\rangle}

\newcommand{\ketbra}[2]{\left|#1\right\rangle\hskip-1mm\left\langle #2\right|}

\begin{document}

\title{Modal, Truly Counterfactual Communication with On-Chip Demonstration Proposal}
\author{Jonte Hance$^{*1,2}$, Will McCutcheon$^1$, Patrick Yard$^1$, and John Rarity$^1$}
\address{$^1$Quantum Engineering Technology Laboratories, University of Bristol, Woodland Road, Bristol, BS8 1US, UK}
\address{$^2$H.H. Wills Physics Laboratories, University of Bristol, Tyndall Avenue, Bristol, BS8 1TL, UK}
\email{*jh14257@bristol.ac.uk}

\begin{abstract} We formalize Salih et al's Counterfactual Communication Protocol (arXiv2018), which allows it not only to be used in with other modes than polarization, but also for interesting extensions (e.g. sending superpositions from Bob to Alice).
\end{abstract}


\section{Introduction}
Counterfactual communication makes use of the amazing property of quantum objects to self-interfere; to change their eventual destination based on not just the path the object has travelled on, but all other paths open to it. To our normal scale, this seems incoherent - as if a person walking, on going down one side of a fork in the road, can still "know" whether the other side leads to a dead end. Despite a series of doubts about the 'counterfactuality' of these schemes, one has finally been created by Salih et al \cite{salih2018laws} that fulfills the three accepted key criteria - a photon arriving at Alice's detectors for both a '0' and '1' result; no weak trace of the photon on Bob's side; and no family of consistent histories placing it at Bob. However, that protocol has the issue of being routed in polarization, making it difficult to implement physically. Here, we abstract it to a modal form, then show how it can then be put into a path-basis, massively easing implementation, and allowing further exploration of it.
\section{Protocol Specifics}
\begin{figure}[htbp]
  \centering
  \includegraphics[width=\linewidth]{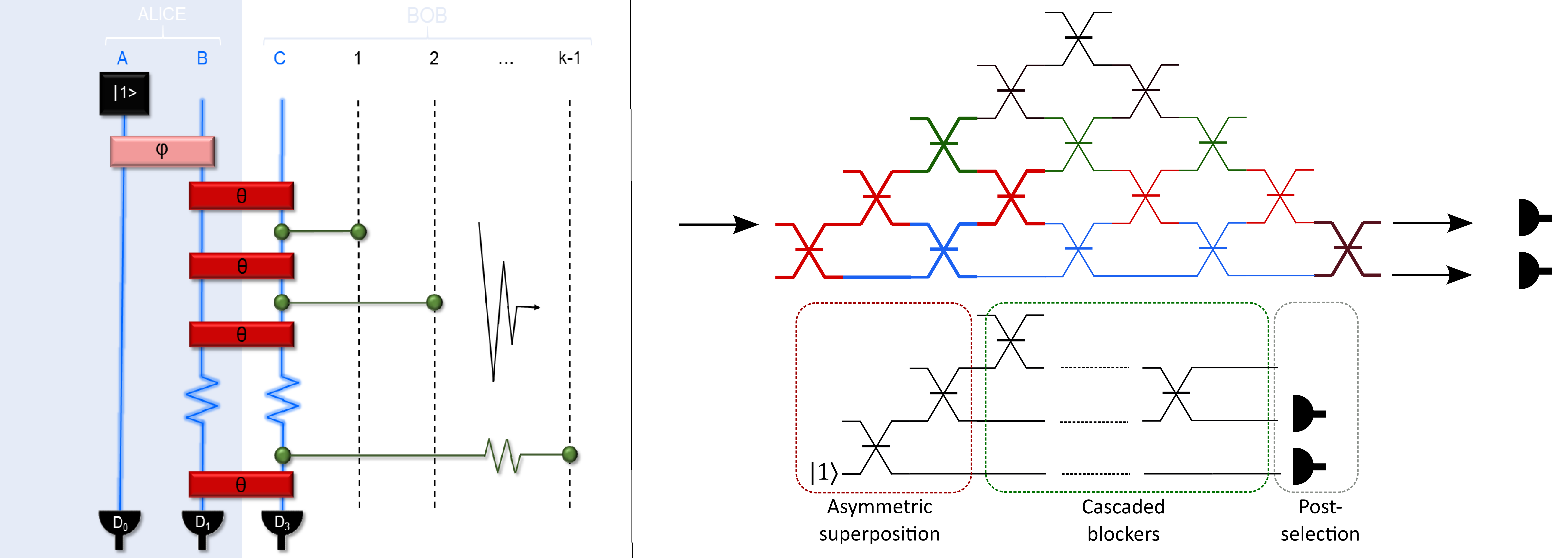}
  \caption{\label{fig:Modal}Diagram giving Hance et al's modal depiction of Salih et al's protocol, and then specifics of how this would be implemented on Carolan et al's Universal Linear Optics chip.}
\end{figure}

We abstract from Salih et al's polarisation-specific form of the protocol \cite{salih2018laws} to the modal one, with three modes, A, B and C, as per Fig. \ref{fig:Modal}.

 Mode A is the channel into which the photon is introduced - with initial state $\ket{A}_{\mathrm{I}}\footnote{Note, for convenience, when dealing with the overall space, we use First Quantization, due to the potentially very large number of Loss Modes under consideration. However, when evaluating subspaces (e.g. that of A \& B), we use the Second Quantization, to allow us to show the possibility of the photon being outside the subspace.}$. The photon ending in mode A causes a detection in the detector $D_0$, corresponding to a bit of value 0 being sent.  Mode B, here, is the path within the inner interferometers that does not go to Bob, with Mode C as the one which does. The photon ending in mode B causes a count at detector $D_1$ (a bit of value 1), and it ending in mode C causes a count in detector $D_3$ (in which case the protocol is aborted and restarted). Further, we have the loss mode channels, $1,2...K$, which represent the loss of the element of the photon from mode C incurred when Bob blocks, for each of the K cycles (inner Mach-Zehnder interferometers) of the protocol.
 
 Onto these modes, we enact various unitaries. The first of these is $\hat U_{AB}^{(\phi)}$, which enacts a rotation of $\phi$ between modes A and B upon the state (HWP1 in Salih et al's protocol), where $\phi = \frac{\pi}{2}-\delta$, for small $\delta$.

The next unitary is $\hat U_{BC}^{(\theta)}$, performing a rotation of $\theta$ between B and C (representing each HWP2 operation between the two). By our set-up, we define $\theta=\pi/2K$.

The third and final unitary used here is $\hat X_{Cn}$, the Pauli $X$ operator, swapping mode $C$ with loss mode $n$ (while keeping A, B, and all other loss modes the same) by
$\hat X_{Cn}=\ketbra{n}{C} +\ketbra{C}{n}+\mathbb{1}_{A,B,\{i\}_{(k, \forall i\neq n)}}$.

Depending on whether Bob is blocking or not blocking mode C, the protocol is represented via operators in one of two ways:

\textbf{Not Blocking:} When Bob is not blocking, the $K$ cycles of the interferometers act as $K$ applications of $\hat U_{BC}^{(\theta)}$. Given the definition of $\theta$, this overall acts as $\hat U_{BC}^{(\frac{\pi}{2})}$ (a rotation of $\frac{\pi}{2}$ between B and C). This operates on $\hat U_{AB}^{(\phi)}$, which acts upon  $\ket{A}_{\mathrm{I}}$, giving
$\ket{NBl_K}_{\mathrm{F}}=\hat U_{BC}^{(\frac{\pi}{2})}U_{AB}^{(\phi)}\ket{A}_{\mathrm{I}}$, where $\ket{NBl_K}_{\mathrm{F}}$ is the final state of the system after a full run of the protocol when Bob doesn't block.

This is always counterfactual as the 0-bit is only recorded when the photon is found in mode A - for which there is now way for it to have been to Bob's blockers and back.

\textbf{Blocking:} When Bob blocks, each of the $K$ cycles acts as the combination of $\hat X_{Cn}$ and $\hat U_{BC}^{(\theta)}$ - first a rotation of $\theta$ from B to C, then a swap between C and loss mode $n$. Therefore, the application of K cycles, acting on $U_{AB}^{(\phi)}\ket{1_A}_{\mathrm{I}} $, is
$\ket{Bl_K}_{\mathrm{F}}=\hat U_{BC}^{(\theta)}(\prod_n^{K-1}\hat X_{Cn}\hat U_{BC}^{(\theta)})U_{AB}^{(\phi)}\ket{A}_{\mathrm{I}}$,
where $\ket{Bl_K}_{\mathrm{F}}$ is the final state of the system after $K$ cycles of the protocol.

This is always counterfactual as, if the photon goes into Bob's channel (C) and encounters the blockers, it is always absorbed by them, meaning a 1-bit, from Mode B, can never have directly encountered Bob.

\section{Performing Protocol On-Chip}
The proposed demonstration of the protocol would be done on Carolan et al's Universal Linear Optics chip.\cite{Carolan711} On such a device, the parameters necessary to allow the chip to operate the protocol can be programmed in easily and quickly, saving the time and cost needed to separately manufacture such a device using integrated silicon photonics, or the massive time and effort needed to undertake such a thing in bulk fibre or free-space.

The blue MZIs in the figure implement an identity, the red MZIs are tuned to allow partial (the first one from the right being tuned to a $\phi$ rotation, and the others to a $\theta$ rotation), and the green MZIs being either tuned to rotate by $\frac{\pi}{2}$ (Blocking) or $0$ (Not Blocking). This illustrates how up to k = 4 can be achieved. By using only the bold MZIs, setting everything else to the identity and the setting the blocker to be a 50:50 beamsplitter it would also be possible to send a superposition state counterfactually. The bottom right MZI (brown) would then be used to perform full state tomography on Alice's qubit.

\section{Conclusion and Outlook}
The work done above allows true counterfactuality to be implemented and examined far more easily - and, by giving a method for sending superpositions, hints at the possibility of entangling photons counterfactually. All these factors make this protocol seem the best way we have for examining the final primitive of quantum mechanics - counterfactuality.

\bibliographystyle{unsrt}
\bibliography{ref.bib}

\end{document}